# Reinterpreting Boltzmann's H-Theorem in the light of Information Theory

DAVID SANDS
Department of Physics and Mathematics
University of Hull
Hull
UK
e-mail: d.sands@hull.ac.uk

JEREMY DUNNING-DAVIES
Institute for Theoretical Physics and Advanced Mathematics (IFM) Einstein-Galilei,
Via Santa Gonda,
14 – 59100 Prato,
Italy

**Abstract**
Prompted by the realisation that the statistical entropy of an ideal gas in the micro-canonical ensemble should not fluctuate or change over time, the meaning of the H-theorem is re-interpreted from the perspective of information theory in which entropy is a measure of uncertainty. We propose that the Maxwellian velocity distribution should more properly be regarded as a limiting distribution which is identical with the distribution across particles in the asymptotic limit of large numbers. In smaller systems, the distribution across particles differs from the limiting distribution and fluctuates. Therefore the entropy can be calculated either from the actual distribution across the particles or from the limiting distribution. The former fluctuates with the distribution but the latter does not. However, only the latter represents uncertainty in the sense implied by information theory by accounting for all possible microstates. We also argue that the Maxwellian probability distribution for the velocity of a single particle should be regarded as a limiting distribution. Therefore the entropy of a single particle is well defined, as is the entropy of an N-particle system, regardless of the microstate. We argue that the meaning of the H-theorem is to reveal the underlying distribution in the limit of large numbers. Computer simulations of a hard-sphere fluid are used to demonstrate the ideas.

## 1. Introduction

Boltzmann's H-theorem rightly stands as a seminal work in statistical physics. It proved theoretically what was initially only conjecture on Maxwell's part, namely that the orthogonal velocity components of an ideal gas were Gauss distributed. Boltzmann saw his theorem as going much further, however. He clearly regarded it as a mechanical proof of the law of increasing entropy, even though there were two cogently argued objections due respectively to Loschmidt and Zermelo. Loschmidt argued that as the laws of mechanics are time-reversible it should be possible to imagine trajectories that do not always increase the entropy and Zermelo argued that Poincaré recurrence would mean that the system would return to its starting state or to some state arbitrarily close to it at some point in the future. Again, entropy would decrease. Boltzmann vigorously defended his ideas [1,2], arguing that as the probability of the system returning to such a state is infinitesimally small, the system can be regarded as having evolved to spend most of its time in the most probable state, which is the state of maximum entropy. The H-theorem is therefore regarded by many physicists as the basis of the idea that the entropy of a closed system evolves to a maximum value at equilibrium.

Interesting though the history and philosophy of these ideas are, of primary interest here is the idea that the entropy of an isolated system remains constant in time. This is clearly at odds with Boltzmann's own interpretation of his H-theorem as well as the view described above regarding the evolution of the entropy of an isolated system, yet it is contained within statistical mechanics. Specifically, it is implied in the connection between a micro-canonical system as described by Boltzmann and a canonical system as described by Gibbs. We describe this contradiction in detail and use computer simulations of a hard-sphere fluid to show how the entropy of a single particle within a classical ideal gas in equilibrium is well defined regardless of the microstate of the system. In consequence, the entropy of the $N$-particle gas is also well defined and constant regardless of the microstate. We offer a re-interpretation the H-theorem in the light of these ideas.

## 2. Canonical and micro-canonical systems

The canonical and micro-canonical distributions become identical in the so-called "thermodynamic limit" of an infinite number of particles, but there is a simpler connection between the two, as described by Gibbs: "If a system of a great number of degrees of freedom is microcanonically distributed in phase, any very small part of it may be regarded as canonically distributed" [3]. Thus, a micro-canonical system can be regarded as a collection of canonically distributed systems. That is, if a system contains $N$ particles and is subdivided into $m$ identical systems, each containing $n$ particles such that $m=N/n \gg 1$, any one of the $m$ systems can be regarded as being canonically distributed with the other $m$-1 systems acting collectively as a reservoir. This links the entropy of the two different distributions because, as described in virtually any text book on statistical mechanics, it is common when deriving the canonical distribution to consider a fluctuation in energy between the reservoir and small system. The corresponding fluctuation in entropy, found by taking a Taylor expansion of the entropy to first order around the energy of the reservoir, leads to the well known expression,

$$p(E_i) = \frac{1}{Z} e^{-\frac{E_i}{kT}} \quad (1)$$

Normally, the effect of the entropy fluctuation on the reservoir is ignored as it is only the small system connected to the reservoir that is of interest, but as the reservoir in this case is made up of the remainder of the micro-canonical system under consideration, the change in entropy of the reservoir becomes important. In fact, as the energy fluctuation is assumed to occur at fixed temperature the entropy change is assumed to be directly proportional to the energy fluctuation. That is, the expansion of the entropy is to first order only, so a fluctuation of $\pm \delta E$ in the energy of the small system, which leads to a change of $\pm \delta S$ in the entropy, is matched by a corresponding fluctuation of $\mp \delta E$ in the energy of the *m*-1 other systems that together comprise the reservoir and a corresponding entropy fluctuation of $\mp \delta S$. Therefore the change in the entropy of the reservoir exactly matches that in the small system and the nett entropy change of the system as a whole must be zero.

If this derivation of the canonical distribution is correct entropy fluctuates between different parts of the system even though the total entropy remains fixed. As is also well known, however, the distribution in equation (1) leads to the following expression for the entropy

$$S = k \ln Z + \frac{\langle E \rangle}{T} \quad (2)$$

with $\langle E \rangle$ being the average energy. There is nothing within the expression for the entropy in (2) that fluctuates, which appears to invalidate the original assumption and with it the derivation. Although this supports the idea expressed by Jaynes over 50 years ago that the canonical distribution is without a firm physical foundation [4], there is no question that the canonical distribution itself remains very important within statistical mechanics. However, if the entropy of a canonically distributed system does not fluctuate along with the energy, but remains constant, it follows that the entropy of a large micro-canonical system does not fluctuate either, regardless of how the energy is distributed throughout the system.

This clearly contradicts the current understanding of the H-theorem [5]. As is well known, the Second Law of Thermodynamics, by which is meant the law of increasing entropy, was seen as having only a statistical validity [6] even before Boltzmann published his H-theorem in 1872, and Boltzmann interpreted his theorem as proof of this law. By his reasoning entropy increases as the system evolves towards the equilibrium (Maxwellian) velocity distribution from an initial non-equilibrium distribution. As we have described, there were notable objections to this interpretation at the time, and in the modern interpretation of entropy as a volume in phase space, it is implied that the system can occupy states of low probability and hence low entropy, even if such events are highly improbable. Thus it seems to be implicit within the interpretation of the H-theorem that the entropy of a micro-canonical system fluctuates. In the remainder of this paper we show how information theory, backed up by computer simulations of hard-sphere fluids, can help resolve this difficulty and lead to a re-interpretation of the H-theorem.

3. **Computer simulations of hard-sphere fluids**

By their nature computer simulations are restricted to small numbers of particles yet the system is describable in terms of the Maxwellian velocity distribution, or, equivalently, the Maxwell-Boltzmann speed distribution, which differs from it by a density of states term and is easier to demonstrate graphically. Consider, for example, one particle of atomic mass 20 among a total of 400 identical particles sampled at regular intervals of time. The particles all start with the same speed but with random positions, which are updated every $10^{-5}$ seconds. After every 150 such updates the velocity of a single particle is sampled. Figure 1 shows the corresponding histogram of speeds from $10^5$ such samples together with the theoretical Maxwell-Boltzmann speed distribution. The motion of this single particle over time is clearly describable by the Maxwell-Boltzmann speed distribution in the sense that the speeds accessible to it, whilst random, are limited by this distribution. It follows that every other particle is also describable by the Maxwell-Boltzmann speed distribution in exactly the same way and the system as a whole must also be describable by the same distribution. Yet, there are only 400 particles and the distribution of speeds over the particles at any given instant will be an imperfect representation of the Maxwell-Boltzmann distribution. Figures 2 and 3 show two such sampled distributions along with the theoretical speed distribution for comparison.

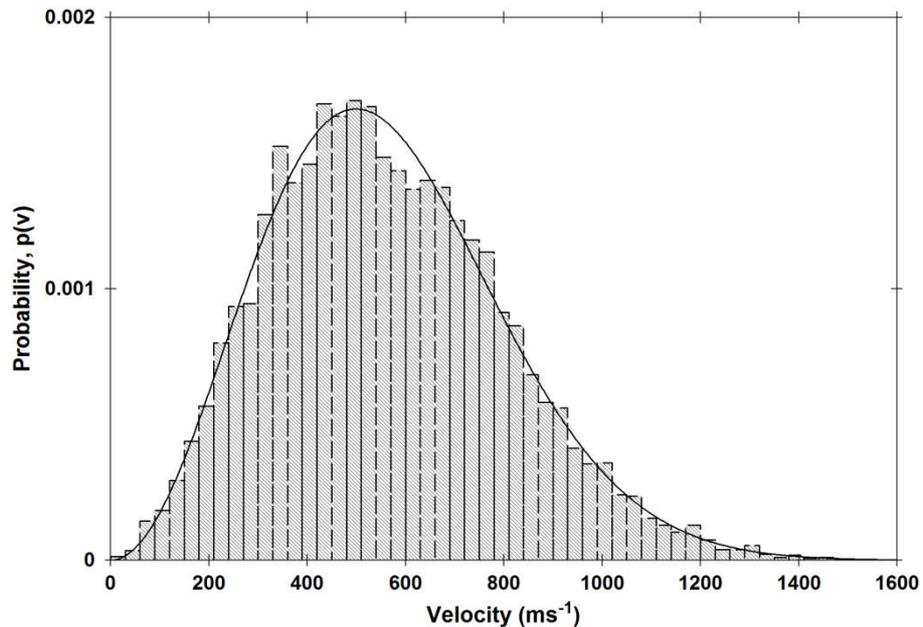

Figure 1. The distribution of speeds for a single particle calculated at intervals of 30 ms$^{-1}$ from a total of $10^5$ samples. The smooth curve is the Maxwell-Boltzmann speed distribution at the temperature of the particles.

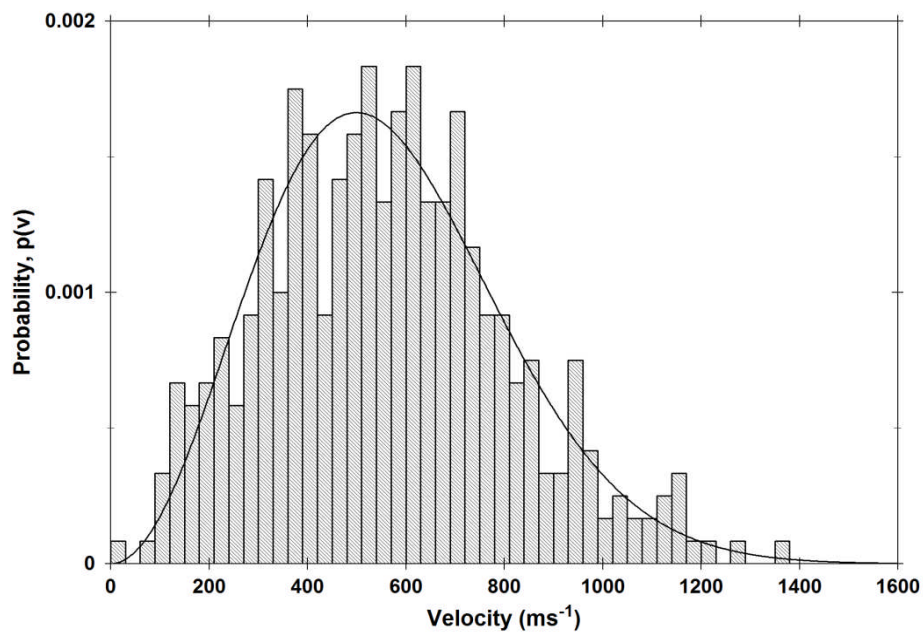

Figure 2. The instantaneous distribution of speeds across all the particles calculated at intervals of 30 ms$^{-1}$ corresponding to the high-entropy sample 10 in figure 4. The smooth curve is the Maxwell-Boltzmann speed distribution at the temperature of the particles.

Although the entropy over the particle distributions can be defined in these circumstances, it will fluctuate with the distribution of velocities across the particles. The distributions in figures 2 and 3 are clearly different, but it is virtually impossible to state, by visual inspection alone, which is closer to the theoretical distribution. However, the distribution in figure 2 yields a higher entropy than that in figure 3, as shown by figure 4, which shows a sequence of 48 different H-functions calculated from the distribution of speeds taken across the particles at time intervals corresponding to 3 seconds or $3\times10^5$ steps of the simulation. Samples 10 and 34 in figure 4 are respectively states of high and low entropy and correspond to figures 2 and 3. Figure 5 shows the average H-function, as well as the standard deviation, calculated over a large number of samples of the distributions of speeds across the particles for systems with up to 2000 particles. As the number of particles increases the average entropy increases but is clearly tending to a limiting value whilst the standard deviation clearly decreases in a like manner. The explanation for this is the following; for large numbers of particles the instantaneous distribution of speeds across the particles corresponds much more closely to the Maxwell-Boltzmann distribution and the likelihood of significant deviations from that distribution decreases. The implication is that in the asymptotic limit of a very large number of particles the distribution across the particles becomes identical to the theoretical Maxwell-Boltzmann distribution.

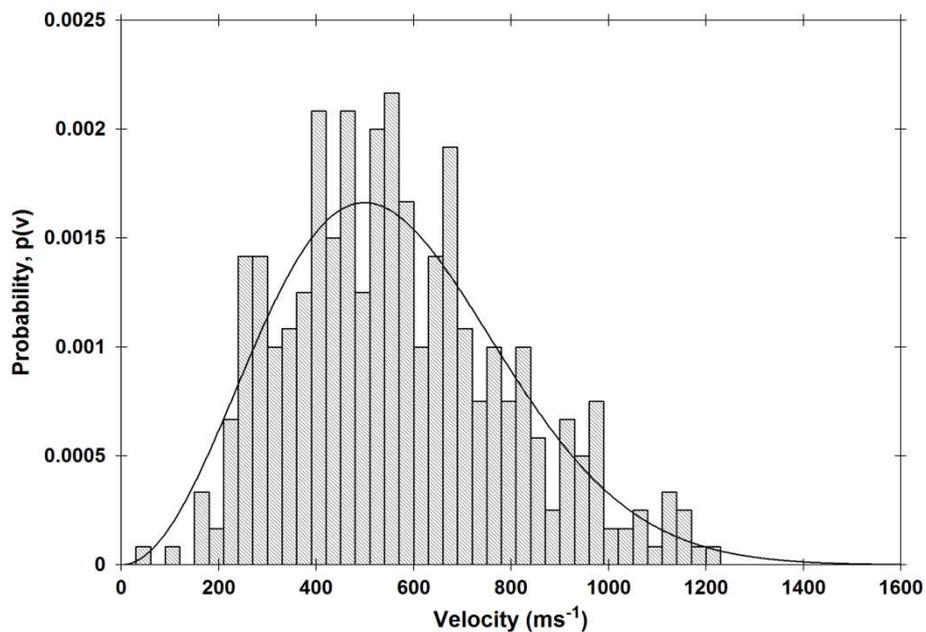

Figure 3. The instantaneous distribution of speeds across all the particles calculated at intervals of 30 ms$^{-1}$ corresponding to the low-entropy sample 34 in figure 4. The smooth curve is the Maxwell-Boltzmann speed distribution at the temperature of the particles.

## 4.    Discussion

We have shown that for a system containing a small number of particles the entropy calculated from the distribution across particles fluctuates with the particular microstate. It should be noted, however, that the entropy shown in figure 4 is not the same as the entropy used in statistical mechanics, which is based on the velocity distribution rather than the speed distribution. However, the two are closely related as the two functions differ only by the inclusion of the density of states. One important difference between statistical mechanics and the present work concerns the dependence of the entropy on the number of particles. Strictly, the entropy of statistical mechanics is taken over the accessible phase space, which increases with the number of particles in the system, but both the velocity and speed distributions are independent of the number of particles. Whilst this is convenient for demonstrating the nature of fluctuations in entropy it makes the relevance to statistical mechanics harder to see.  For this reason we concentrate on the distribution over velocity states for a single particle. We include also the spatial distribution of the particle in the entropy in order to make direct comparison with statistical mechanics. Finally, having discussed the entropy of a canonical system we shall return to the Maxwellian distribution over the particles in order to re-interpret the H-theorem.

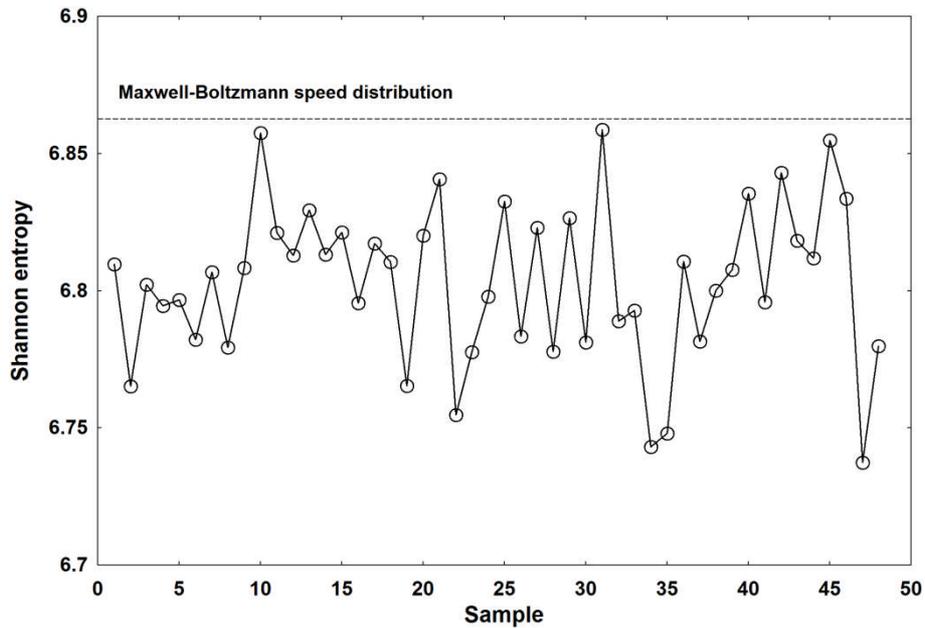

Figure 4. The Shannon entropy (H-function) calculated from 48 different instantaneous distributions of speeds across all the particles calculated at intervals of 30 ms$^{-1}$. The dotted line corresponds to the entropy calculated from the Maxwell-Boltzmann speed distribution at the temperature of the particles.

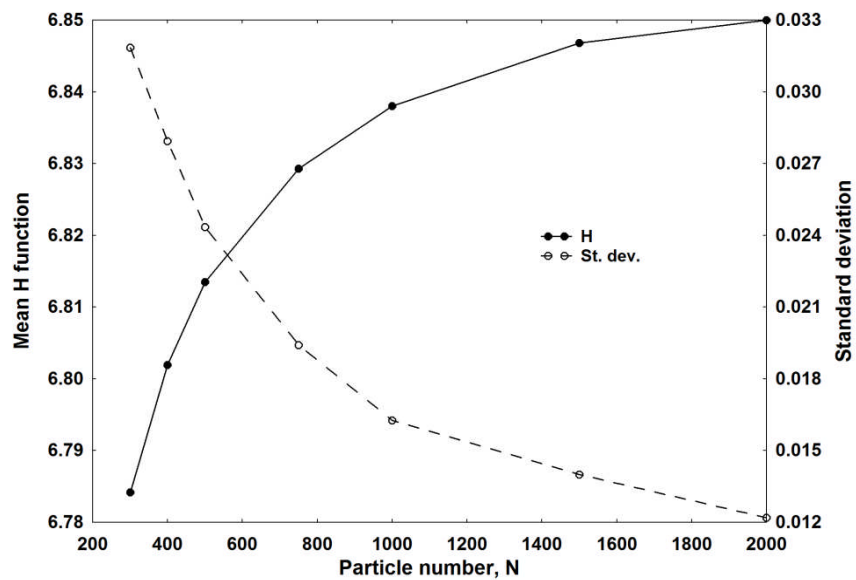

Figure 5. The mean H function, together with its standard deviation, as a function of the number of particles in the system.

The expression that Boltzmann derived for the entropy is the same as that later derived by Shannon in the context of what has now come to be known as information theory [7] and it should be possible to interpret it in the same way as a measure of uncertainty. That is, at any given time both the velocity and position are unknown and must therefore be described by probability distributions. If the volume of the system is $V_S$ the entropy associated with a single particle is

$$S_1 = -k\left[\int_0^\infty p(v)\ln[p(v)]dv - \frac{1}{V_S}\ln[V_S]\int_0^{V_S} dV\right] \quad (3)$$

It has been shown [8] that if $p(v)$ is given by the Maxwellian velocity distribution this leads to the well known expression

$$S_1 = \frac{3}{2}k\ln T + k\ln V_S + terms \quad (4)$$

For a canonical system as described in the introduction, each particle has access to the full range of velocities defined by the Maxwellian distribution. Accordingly, the probability that the system occupies a volume of phase space $d\tau$ centred on a phase $\tau$ is simply an *n*-fold product of single-particle probabilities [9]. That is, if $\mu$ represents the coordinates of a single particle in phase space then;

$$P(\tau)d\tau = p_1(\mu_1)d\mu_1 \cdot p_2(\mu_2)d\mu_2 \cdot p_3(\mu_3)d\mu_3 \ldots p_n(\mu_n)d\mu_n = p(\mu)^n d^n\mu \quad (5)$$

It follows that the entropy of the *n*-particle canonical system is

$$S_n = nS_1 = \frac{3}{2}nk\ln T + nk\ln V_S + terms \quad (6)$$

Central to any further discussion are the notions of a single-particle Maxwellian and the corresponding single-particle entropy. The latter is defined because a probability distribution can be associated with a single particle, but clearly a single particle cannot have access simultaneously to all possible states. There is therefore a crucial difference between the Maxwellian distribution applied to a system and the same distribution applied to a single particle; the former gives the expected number of particles in a given velocity state, by which is meant here a given interval, at any given instant in time whilst the latter gives the expected number of times the particle has occupied, or will occupy, a given state. This is shown in figure 1. The single particle speed distribution clearly corresponds to Maxwell-Boltzmann. It is noticeable, however, that even after $10^5$ samples of the particle's speed the observed probabilities differ from their expected values. This implies that if the observed probability distribution is used in equation (3) the value of the entropy will depend on how many observations are made, assuming such observations to be possible. Not only will the entropy fluctuate if repeated observations are made but the value will steadily approach a limit the longer the particle is observed. Physically this makes no sense.

As we have described, entropy in information theory represents uncertainty; that is to say, the lack of information about a system means that it is not possible to predict the outcome of a particular event. If this were not the case and the outcome could be specified with complete certainty beforehand, there would be no need to invoke probability and the entropy would be zero. Applied to the single particle, both the velocity and position are uncertain. We can do no more than specify the probability that either will be found within a certain range at any given time and the information entropy is a measure of this uncertainty. The probability distributions in equation (3) are therefore independent of the observed occupancies and must instead correspond to the limiting, or theoretical, probability distributions.

This is the essential conclusion; the entropy of *any* particle in an ideal gas at a fixed temperature is well defined, depends only on the temperature of the gas and the volume occupied, is independent of the actual state of the particle, and does not depend on any observations that might be made on the particle. Equations (5) and (6) then show that the entropy of a canonically distributed $n$-particle sub-set of the system is also well defined, regardless of the state of the particles. As a micro-canonical system can be regarded as a collection of canonically distributed small systems it then follows, in agreement with equation (2), that the total entropy of the system is also well defined and independent of the particular micro-state.

This brings us to the interpretation of the H-theorem. As described in a recent paper by Badino [10], who has re-examined the history of Boltzmann's H-theorem, Boltzmann himself regarded the H-theorem as a rigorous probabilistic law because the H-theorem was "set up within a framework of proper asymptotic conditions", which includes the limit of very large numbers of particles. Badino is essentially concerned with correcting the common view that Boltzmann's own view of his theorem changed from being a mechanical description of the evolution of a system to a probabilistic view. He argues that although Boltzmann's language is deterministic his view of the H-theorem was probabilistic. However, even though Boltzmann recognized the importance of asymptotically large numbers to a statistical, or probabilistic, argument it would appear that he did not recognize the full implications of a limiting distribution.

As pointed out by Uffink [11], the idea of a limiting distribution is perhaps best regarded as an informal view derived from experience rather than a formal consequence of the law of large numbers. We have argued that the entropy as understood within information theory is a property of the limiting distribution. We have concentrated on the probability distribution associated with each individual particle, but the extension to the distribution across particles is straightforward. As figures 2 and 3 show, the actual occupancy of any given velocity state differs from the expected value for small systems and the distribution is therefore an imperfect representation of the limiting distribution. The entropy calculated from the actual distribution will fluctuate with the microstate of the system but the entropy calculated from the limiting distribution will not. Taking information theoretic entropy as a measure of uncertainty, again the occurrence of all possible combinations, or outcomes, should be accounted for in the probability distribution upon which the entropy is based. This implies that the limiting distribution should be used for the entropy.

It is possible to regard the initial state of uniform speed but random direction used by Boltzmann as simply a microstate of the final distribution which must be accessible at some time in the future, however unlikely the event. This is essentially the recurrence argument of Zermelo and the time-reversibility argument of Loschmidt; trajectories exist which can return the system to the initial state. Both objections are essentially correct and dispel the notion of the system evolving to a final state as there is no final state in a finite system, which will be in a constant state of flux. Regardless of the microstate, however, the system is always describable in terms of the limiting distribution, both in terms of the likelihood of observing a particular velocity state for an individual particle and the likelihood of observing a given distribution of velocities, or speeds, across all the particles.

Correct though this argument undoubtedly is for small systems such as we have examined through computer simulations, it would not appear to be valid in the asymptotic limit of large numbers of particles when the probability of a significant deviation from the theoretical distribution is either negligibly small or zero. Perhaps Boltzmann was seduced by the notion of having discovered a proof Clausius' extension of the Second law, but he appears not to have recognized the contradiction inherent in his interpretation; the infinitesimally small probability of observing a so-called "low entropy", and therefore highly improbable state, cannot be used as a justification or defence of the argument that entropy is maximised over time [1, 2] because if the evolution toward a state of maximum entropy is supposed to occur in real physical systems, the system has to start in such an improbable state. Yet, how does it arrive in such a state in the first place? In the kind of transformation considered by Boltzmann the macroscopic variables of state, such as temperature, pressure and volume, do not vary so this is quite unlike the irreversible processes considered by Clausius. There are no external parameters or constraints that can be adjusted to bring about the sort of conditions that Boltzmann assumed as the starting state.

In the sense that the assumption of an arbitrary starting state is not a reflection of the behaviour of a real physical system, either because in the limit of large numbers the state is effectively inaccessible or because for small systems the state can be revisited over time and is therefore not a starting state in any meaningful sense, it should perhaps be seen as an artificial device. In the asymptotic limit of large numbers as used by Boltzmann, the distribution of velocities across the particles essentially corresponds to the limiting distribution. The assumption of an arbitrary distribution is equivalent, therefore, to the assumption of an arbitrary limiting distribution. If this distribution is stable against collisions it represents the equilibrium distribution. If not, minimising H reveals the limiting distribution. However, there is no sense in which the limiting distribution changes over time as our simulations show that the system is always describable by the Maxwell-Boltzmann distribution. Taking into account the meaning of information entropy and the evidence from statistical mechanics that entropy is constant over time, the meaning of the H-theorem, therefore, is not to show that a system evolves to a state of maximum entropy, but simply to reveal the limiting distribution.

## 5. Conclusion

Computer simulations of a hard-sphere fluid have been used to investigate the apparent contradiction between the idea that statistical entropy should be constant and the idea in Boltzmann's H-theorem that entropy increases over time. As in the derivation of the H-theorem, the simulations start with the particles all having the same speed and sampling of the speed distribution across the particles some time later shows clearly that it corresponds to Maxwell-Boltzmann. Unlike Boltmann, however, we can consider only small numbers of particles. With only 400 particles the distribution of speeds across the particles fluctuates and so does the entropy calculated from it. A small system such as this brings out the nature of the reversibility objections, which are really concerned with the existence of trajectories that cause the distribution to fluctuate. Information theoretic entropy, on the other hand, represents uncertainty and should not fluctuate. This suggests that the distribution used to calculate the entropy should be the limiting distribution. This applies both to the distribution of velocities across the particles as well as the velocities accessible to a single particle over time. Thus the speed of a single particle has a well defined entropy and the entropy of the whole system is similarly well defined. Therefore the meaning of the H-theorem lies not in demonstrating or proving the Second Law but in revealing the limiting distribution in the asymptotic limit of large numbers.